# Annotations for Supporting Collaboration through Artefacts

*Syavash Nobarany*

*March 8, 2011*

*Updated on November 7, 2012*

## Introduction

Shared artefacts and environments play a prominent role in shaping the collaboration between their users. This article describes this role and explains how annotations can provide a bridge between direct communication and collaboration through artefacts. The various functions of annotations are discussed through examples that represent some of the important trends in annotation research. Ultimately, some of the research issues are briefly discussed, followed by my perspective on the future of asynchronous distributed collaborative systems with respect to annotations.

## Collaboration through Artefacts

In the physical world, collaboration in performing physical tasks is always mediated by the physical objects involved in them. For example, when two people carry a table, their movements are transferred through the table, hence, they do not need to talk about coordinating each movement [16]. However, this example describes a synchronous co-located collaboration. Collaboration through artefacts in asynchronous mode requires persistence of the effects that each collaborator may have on the shared artefact or environment. We can observe this kind of collaboration in people's social life as well as other species. For example, ants mark their path by pheromone so that other ants find the way to the food, and dogs mark their territory with urine. Whenever asynchronous coordination between agents is required, they modify their shared environment to communicate with others. This is called stigmergy which was first used to describe how termites' individual behaviour can lead to collective complex behaviour [5].

Direct collaboration in distributed settings is not possible in a purely physical world because the agents cannot sense each other's effects as they do not share the same environment. In the digital world, virtual co-location enables physically distributed people to collaborate in multipe virtual spaces; however monitoring multiple environments is challenging and the design of these virtual environments should help users maintain their awareness of them.

## Annotations in Collaborative Work

Collaboration in asynchronous distributed settings can be mediated through artefacts or can be based on explicit messages between collaborators [4]. Annotations can provide a bridge between the two by enriching the first one and putting the second one in context. The two methods can be considered as the two ends of a spectrum, in which direct communication supported by artefacts, and artefacts enriched by communicative artefacts (in other words, annotations) fall in between (figure 1). This perspective extends the concept of annotation beyond the traditional view (e.g. defined by Webster as 'adding note') and provides a more coherent view of communication in collaborative systems.

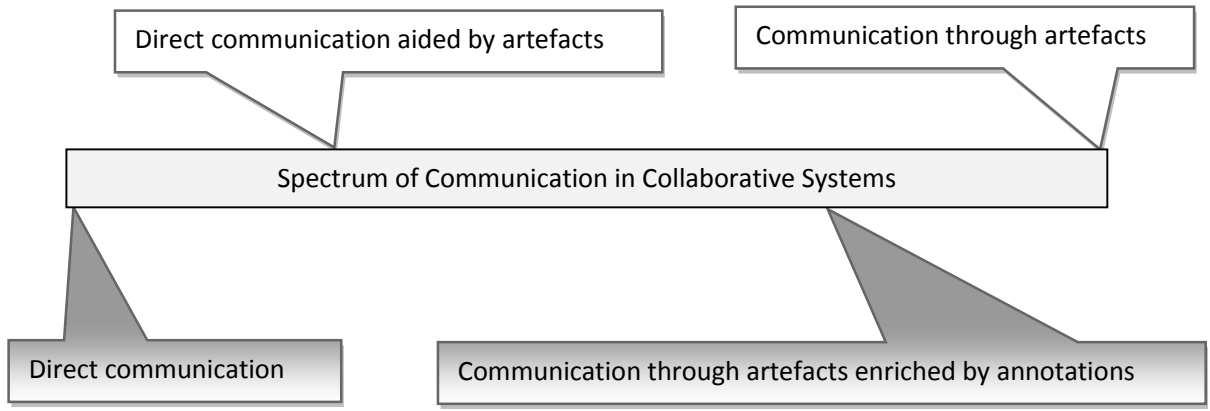

**Figure 1 Spectrum of Communication in Collaborative Systems**

Annotation is a broad concept with many different functions such as managing attention, organization, record of interpretation, summarization, communication, etc. Previous works described various dimensions of annotations [11] and even tried to formalize this concept to provide a unified model of annotations [1]. Although such models can help make sense of the design space of annotations, they do not fit well within the scope of the current work. I focus on collaborative functions of annotations and representative research trends related to them. Based on reviewing a variety of collaborative systems that take advantage of annotations, I identified three salient themes of collaborative functions of annotations: coordinating activities around an artefact, facilitating the discovery of shared artefacts, and supporting comprehension of shared artefacts. In the following sections each of these functions are described and illustrated with examples form recent Computer Supported Cooperative Work (CSCW) research.

## *Coordinating Activity around a Shared Artefact*

The modifications made to an artefact by a collaborator, provides clues to other collaborators about the state of the work and the activities that are needed for advancing the collaborative task. However, these clues are usually not sufficiently expressive, and the collaborator who has made the modifications, needs to communicate further details about the process including, how the modifications have contributed to the progress, what others need to know about the various aspects of modifications that are not visible, and how others may contribute to advance the task. The communication between collaborators can best happen in the context of shared artefacts, to help establish common ground and convey artefact-related messages. Collaborative authoring tools can provide various forms of annotations, to support the coordination of the authoring process. Annotations on a shared document can help attract attention of collaborators to specific parts of the document, make the trace of collaborators visible, and allow for verbal discussions for coordination whenever required.

Cadiz, Gupta and Grudin [3] investigated the use of Word 2000 web annotations by program managers, developers and testers for developing software specification documents. Based on interviews with ten of the 450 annotators, they identified six factors that influenced the usage of annotations. Maintaining the link between annotations and their corresponding artefacts turned out to be the most important reason for the users who stopped using the system. Maintaining awareness of changes, slow pace of communication through annotations, lack of richness, and need for different levels of communication privacy were other influential factors in using Word annotations.

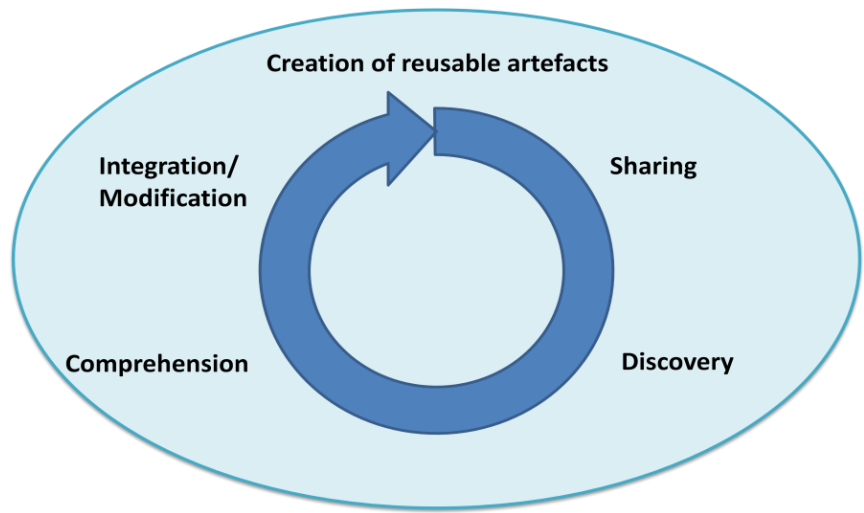

**Figure 2 Cycle of Reuse, adapted from Sumner and Dawe, 2001**

Another telling example of using annotations for coordinating activities is the article development workflow management in Wikipedia. Wikipedia contributors, use annotations to determine the tasks that are needed for improving quality of articles. For example, adding "citation needed" notes within articles, and adding standard notes on top of the articles, that mention the needs of the articles are some of the methods of shaping the workflow of developing articles. Despite several studies on understanding implicit and explicit coordination mechanisms in Wikipedia (e.g. [9,20]), to my knowledge the role of annotations embedded in articles is not yet sufficiently investigated.

*Facilitating the discovery of shared artefacts*
In large collaborative environments, artefacts generated by users may become an input for other users to perform their tasks and to build new artefacts based on them. Sumner and Dawe identify a cycle of reuse by observing the reuse process in a scientific community [19]. The reuse cycle (figure 2) starts with creating an artefact and sharing it, and continues with discovering the resource by other users, understanding it, integrating the artefact into a new task and ultimately sharing the outcome as a new resource. This discovery of artefacts by other users of the environment is an import step that can be facilitated by annotations.

Collaborative tagging systems facilitate the discovery of artefacts, though collecting metadata in the form of keywords to describe shared artefacts. A study of tagging behaviour in Del.icio.us, a collaborative bookmark sharing system, has shown that tags are used to identify artefact's contents, related topics, related tasks, its owner, and its characteristics [6]. A study of tagging behaviour in Movielens, a movie recommender system, showed that people use factual tags (e.g. Action, Drama, Disney, etc.), subjective tags (e.g. overrated, funny, etc.) and personal tags (e.g. netflixQ, buy, etc.) to describe movies [15]. Moreover, factual tags can best support discovering movies while all types of tags (especially personal tags) are helpful for organizing movies. Tagging has also been used to facilitate navigation of software code. Software developers can use TagSEA [18] to associate tags with parts of the source code and use the tags for communication about the code, for example in code reviews, or for guiding newcomers to a software project.

*Supporting Comprehension of Shared Artefacts*

A major drawback of collaboration purely through artefacts is that modifications made to the environment or shared artefacts, are often not rich enough to convey their meaning, the purpose behind them, and the expectations of other collaborators associated with them. Therefore, a major function of annotations is enriching the collaboration through artefacts by explaining activities around artefacts, summarizing them, and enabling other collaborators to understand them and coordinate their activities. This is especially important in the domains where artefacts or changes in the environment are complex. For example, in collaborative software development, comments within the code play a prominent role, mainly because the code developed by a developer is not easily understandable by his collaborators. CodeTalk [17] is a recent work in this area that enables marking and putting comments (similar to word processing tools) on software code. The preliminary evaluations showed that developers found it helpful for informal communication about software code.

Another application domain for using annotations to facilitate comprehension is collaborative data analysis. Sense.us [7] is a collaborative visual analytics system that provides various annotation tools to enable analysts communicate their analysis and their understanding of data. An exploratory evaluation of Sense.us showed that the annotations were mainly used to share understandings and hypotheses, and to ask questions about data.

# Research Issues

In this section, some of the common and important technical and conceptual issues related to annotations are briefly discussed.

*Linking and Orphaning*

Often annotations are associated with specific parts of artefacts. When a collaborator modifies an artefact, correctly repositioning and linking annotations to the corresponding part of the artefact may become challenging. Disconnection of an annotation from its corresponding part of the artefact is referred to as orphaning. The study of Word 2000 web annotations [3] identified orphaning as a key problem. A follow-up work analyzed various aspects of text that can be used in repositioning annotations, and improved their algorithm by searching for sections of the annotated paragraphs [2]. Although more complex algorithms could alleviate the problem, solving the problem in complex situations and for dynamic data representations is still a challenge [2,7].

*Awareness and Privacy*

Maintaining awareness of changes made to shared artefacts or environments is another common issue in distributed collaborative systems. In large collaborative systems, it is important to support unanticipated reuse of artefacts [10,13]. It is hard to know which artefacts and which annotations should be shared with others, especially when people are not aware of how their contributions may be helpful to others [11].

Annotations may reveal personal views or contain pointers to private information [11]. From the perspective of the one who creates annotations or artefacts, it is desirable to share contextual information to ensure the effectiveness of communication; however, it may threaten his/her privacy. This requires managing the trade-off between privacy and awareness [14]. This trade-off is especially important when dealing with tacit annotations that are automatically generated when using shared artefacts such as usage traces [11,12].

*Disturbance and Stifling Creativity*

Not all annotations are interesting or useful for other users of annotated shared artefacts. Particularly, when a shared artefact is annotated by several users, seeing all annotations can be bothersome, even if they are relevant [8]. For example, a large number of visitors of a popular website may want to share their opinions; however, annotations of one or two readers may be sufficient to disturb other readers. Consequently, filtering of annotations is required to manage the complexity of the annotated shared environments [8].

Another possible challenge caused by sharing of annotations is stifling creativity, especially when a shared artefact has several aspects and the annotations focus on a specific aspect and a limited interpretation of the artefact. For example, this may happen in collaborative data analysis systems. Annotations guide the attention of other users to specific interpretations of data [7], which may inhibit fully understanding various aspects and interpretations of the data. To my knowledge, this issue has not yet been investigated and further research is needed in this area.

## Conclusions and Future Directions

Based on my explorations of the design space of asynchronous distributed collaborative systems, I observe and advocate a trend toward scalable, flexible and open collaborative workflows in which people take different roles depending on the needs of the system at different times. The tasks are broken down into very small components, to facilitate micro contributions. Wikipedia is an example of converting a disciplined process of developing an encyclopedia, to an open large-scale collaborative effort. Another example is tagging of resources in collaborative tagging systems, which used to be performed through specific indexing processes by librarians [6].

The role of annotations in such collaborative systems is to provide light-weight implicit and explicit coordination mechanisms that enable people to effectively and efficiently define tasks and micro-tasks, find the ones that they can contribute to, and communicate about them.

The role of collaborative system designers is to design platforms that enable open, flexible, and scalable collaborative workflows, and to constantly refine the platforms to address the evolving needs of the communities running them.

Finally, CSCW researchers need to analyze these workflows and discover mechanisms that enable and facilitate smooth transition to them, especially in complex and sensitive domains where high quality and reliability are required.

## Acknowledgement

I thank Joanna McGrenere for her helpful feedback on an earlier draft of this article.